\documentclass[a4paper,11pt]{article}

\usepackage{pos}
\usepackage{hyperref}

\usepackage{etoolbox}
\title{The SST-1M stereoscopic system}
\author*[a]{C.~Alispach}
\author[a]{T.~Montaruli}

\affiliation[a]{\textit{DPNC - Universit\'e de Gen\`eve, 24 Quai Ernest Ansermet, CH-1211 Gen\`eve,  Switzerland}}

\onbehalf{for the SST-1M Collaboration}

\emailAdd{cyril.alispach@unige.ch}

\abstract{The Single-Mirror Small-Size Telescope (SST-1M) is an Imaging Atmospheric Cherenkov Telescope designed for detecting very high-energy gamma rays. With a compact design achieved through the adoption of silicon-photomultiplier pixels and a lightweight structure, SST-1M offers a large field of view of about 9° and features a mirror system of 4 m diameter with an optical PSF (at 80\% of photon inclusion) of 0.08° on axis and 0.21° at 4° off-axis, and a fully digitizing readout almost deadtime free up to few kHz. The SST-1M achieved a high-performance and cost-effective solution for implementing an array of small-sized telescopes.
The stereoscopic system of two SST-1Ms is temporarily installed at the Ond\v{r}ejov Observatory in the Czech Republic. From an altitude of only about 510 m and in harsh meteorological conditions, the system is detecting galactic sources and flares of AGNs. The accurate calibration of the detector and the simulation benchmark are ongoing. The results of its performance are shown. A future final location is being considered and a future performance outlook is discussed.}

\ConferenceLogo{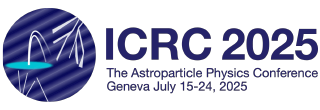}

\FullConference{39th International Cosmic Ray Conference (ICRC2025)\\
 15–24 July 2025\\
Geneva, Switzerland\\}

\begin{document}
\maketitle

\section*{Introduction}

The Small-Sized single mirror Telescopes (SST-1Ms)\cite{Alispach:2024gvp} have been developed for high-energy gamma-ray astrophysics. The SST-1M design was originally proposed as a possible implementation of the 70 telescope array of small sized telescopes of the Cherenkov Telescope Array Observatory (CTAO). They are designed to be a cost-effective, robust, and high-performance solution for detecting gamma-rays in the TeV energy range in an array. A description of the SST-1M telescope design is given in section~\ref{sec:sst-1m}.

Several calibration campaigns of the instruments have been undertaken during commissioning (see \cite{Alispach:2024gvp}) and during operation (see \cite{ICRC25Operation,CrabAA}) of the telescopes and are highlighted in section~\ref{sec:calib}.

The two telescopes were installed, 152.5~m apart, in 2022 at the Ond\v{r}ejov Observatory in Czech Republic at 510~m a.s.l. and conduct stereoscopic observations since October 2023. Observation campaign of several gamma-ray sources are summarized in section~\ref{sec:obs} and are further discussed in these proceedings \cite{ICRC25CTA1, ICRC25DragonFly, ICRC25Mrk421, ICRC25Crab, CrabAA}.

Finally, in section~\ref{sec:future}, the potential next location and potential for high-energy astrophysics of the SST-1M telescopes will be discussed. The potential of a hybrid detector array with the SWGO is discussed in these proceedings \cite{ICRC25Hybrid}.

\section{The SST-1M telescopes}\label{sec:sst-1m}



\subsection{Telescopes Design and structure}

\begin{figure}
    \centering
    \includegraphics[width=0.49\linewidth]{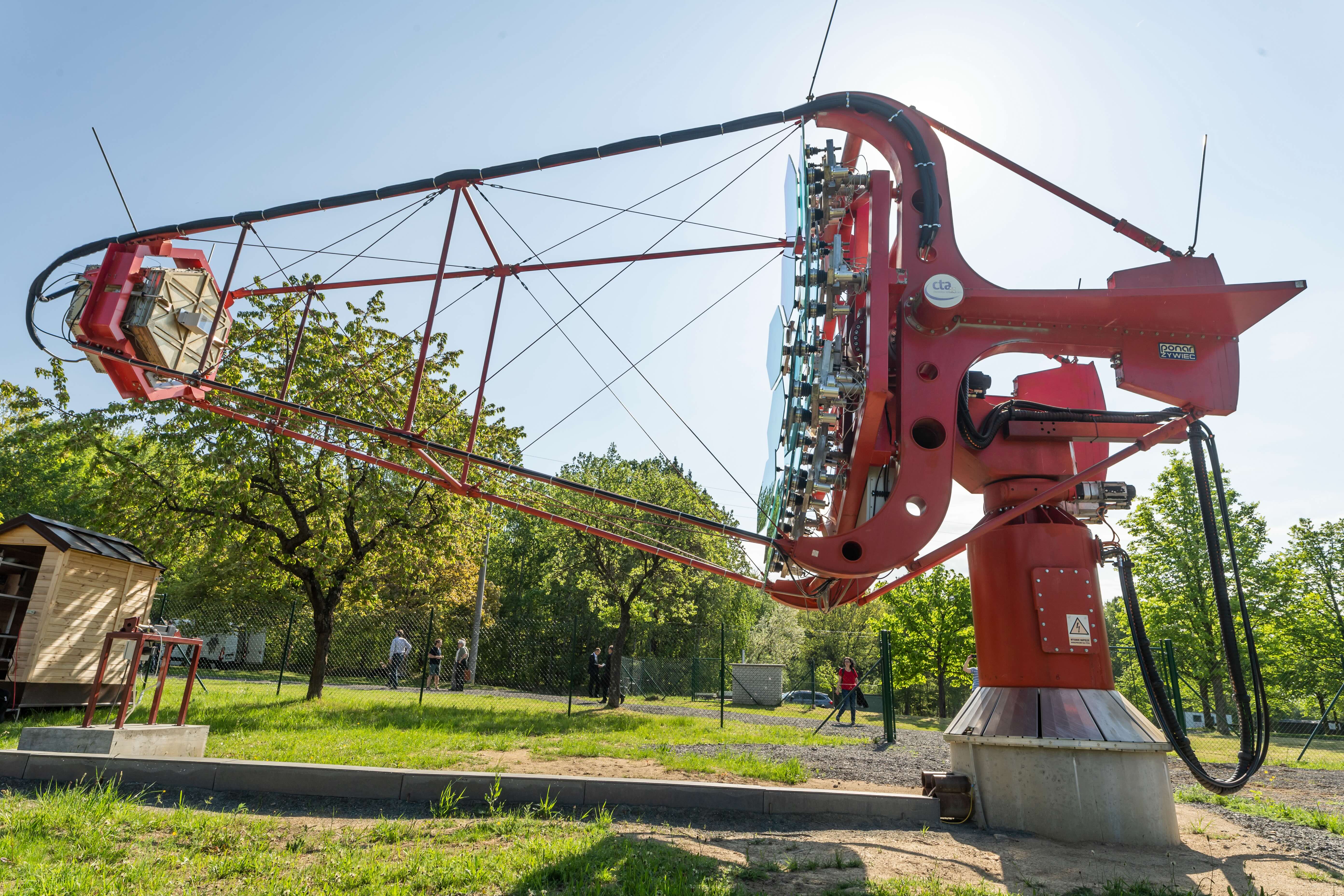}
    \includegraphics[width=0.49\linewidth]{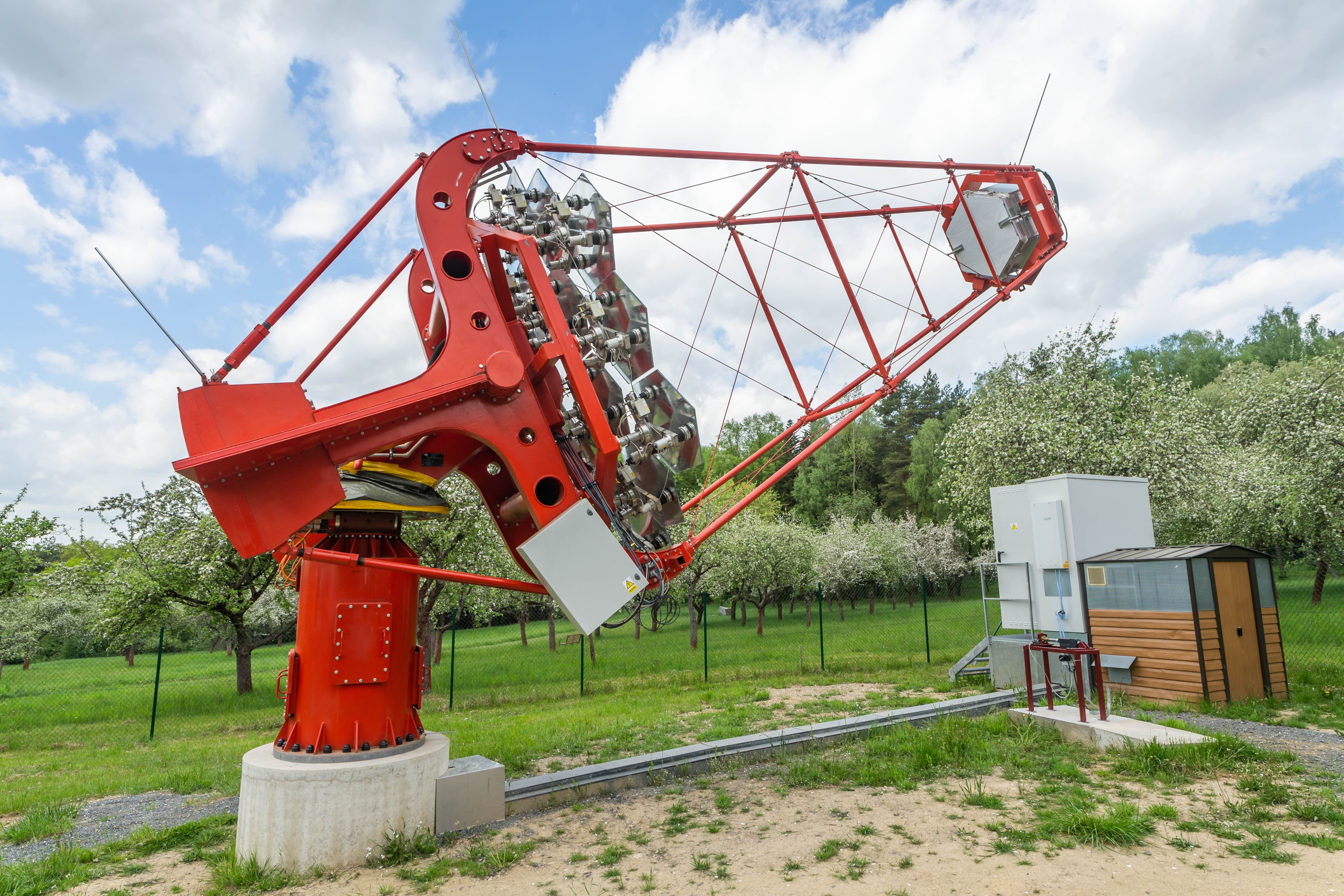}
    \caption{Images of the SST-1M-1 (left) and SST-1M-2 (right) telescopes at the Ond\v{r}ejov Observatory in Czech Republic 510~m a.s.l.. The telescopes are placed 152.5~m apart and work jointly as a stereoscopic system. Taken from~\cite{Alispach:2024gvp}.}
    \label{fig:telescopes}
\end{figure}

The SST-1M telescopes (see Figure~\ref{fig:telescopes}) feature a compact and lightweight mechanical design, with a total individual mass of approximately 8.6 tons. The structure is primarily fabricated from standard steel profiles and tubes, which simplifies manufacturing and reduces costs. Its modular design allows each entire telescope to be transported in a standard 12-meter shipping container\cite{Alispach:2024gvp}. 

The telescope is mounted on a reinforced concrete foundation and includes robust support systems comprising a tower and a rotating head. The movement of the telescopes is controlled via two independent drives for azimuth and elevation. The system supports rapid slewing to any sky position within one minute and achieves a pointing accuracy of 7 arcseconds and a tracking accuracy of 5 arcminutes~\cite{Alispach:2024gvp}.


\subsection{Optical systems}

The optical system of the SST-1M is based on a Davies-Cotton (D-C) configuration, optimized for wide 9° field-of-view imaging. It consists of 18 hexagonal mirror facets, each 78 cm across, arranged in two concentric rings to form a 4-meter diameter spherical dish for an effective mirror area of 6.47~m$^2$. The mirrors are made of 15 mm thick borosilicate glass and are coated with aluminum and a protective silicon dioxide layer to maximize reflectivity in the 300–550~nm wavelength range, which matches well to the Cherenkov light spectrum of extensive air showers.

\begin{figure}
    \centering
    \includegraphics[width=0.6\linewidth]{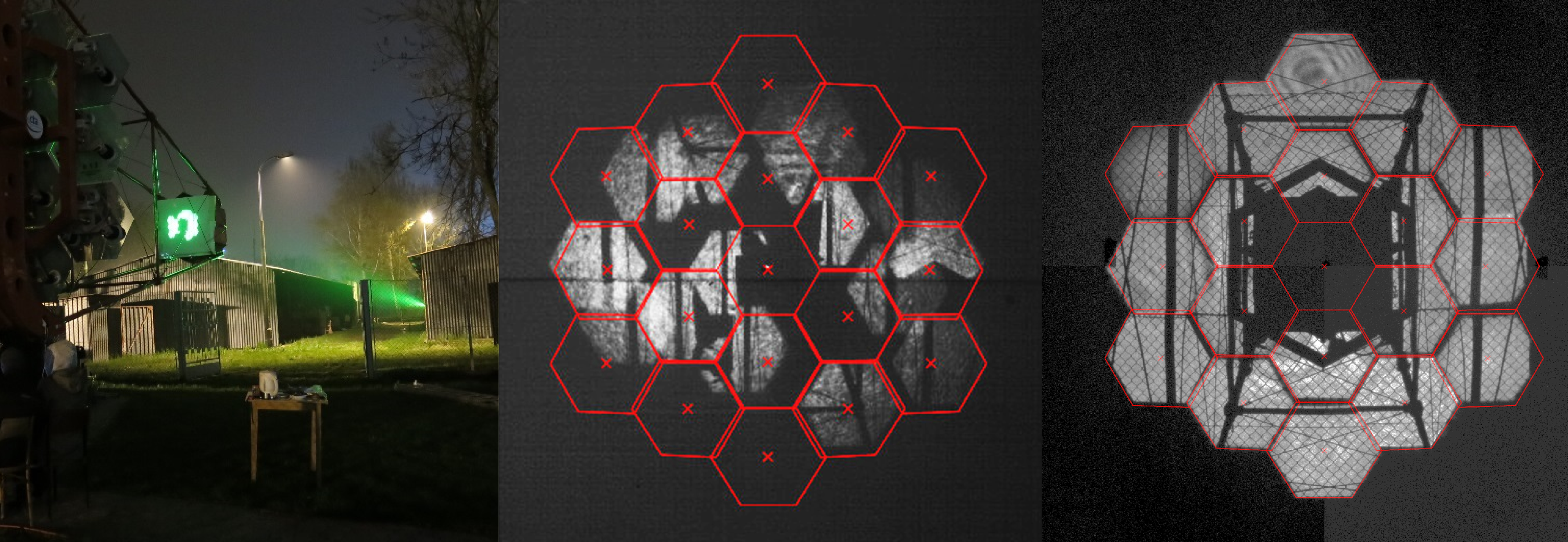}
     \includegraphics[width=0.35\linewidth]{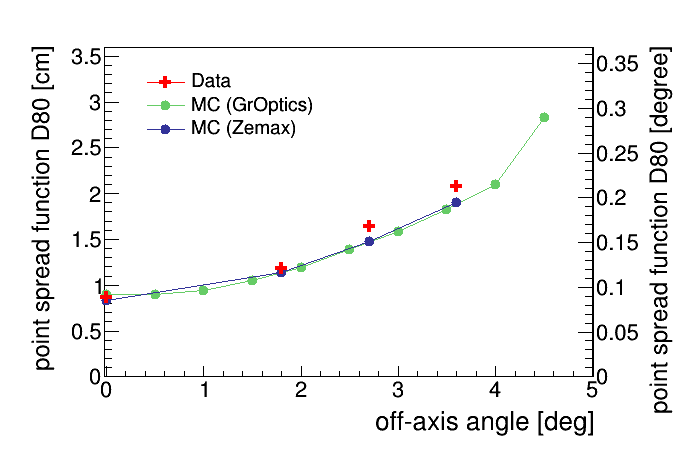}
    \caption{\textbf{Left:} Photograph of the mirror alignment procedure. A point-like laser source, positioned approximately 30 meters from the telescope, illuminates a screen placed in front of the camera. The projection of the source, is used to adjust each mirror facet so that its reflection aligns with a predefined target location on the screen. \textbf{Center:} Images of the mirror facet reflections before and after the alignment process, respectively. \textbf{Right:} Measured (red) point spread function (80\% containment diameter) as function of the off-axis angle compared to ray-tracing Monte Carlo simulations (green and blue). Taken from~\cite{Alispach:2024gvp}.}
    \label{fig:psf}
\end{figure}

Each mirror is mounted and aligned with two actuators and a fixed point, allowing for fine adjustments and thermal decoupling. The alignment process is supported by a dual-camera system: one CCD camera monitors the star field reflected on the telescope lid for astrometric calibration, while a second camera observes the point spread function (PSF) during alignment and operation. This setup ensures that the optical performance remains within specifications under varying environmental conditions. Figure~\ref{fig:psf} illustrates the mirror alignment procedure performed with a laser source and the measured PSF as function of the off-axis angle ranging from 1° (on-axis) to 3° (at the edge of the camera). The measurements match well the ray-tracing Monte Carlo simulations. 

\subsection{Control software and operation}

\begin{figure}
    \centering
    \includegraphics[width=0.3\linewidth]{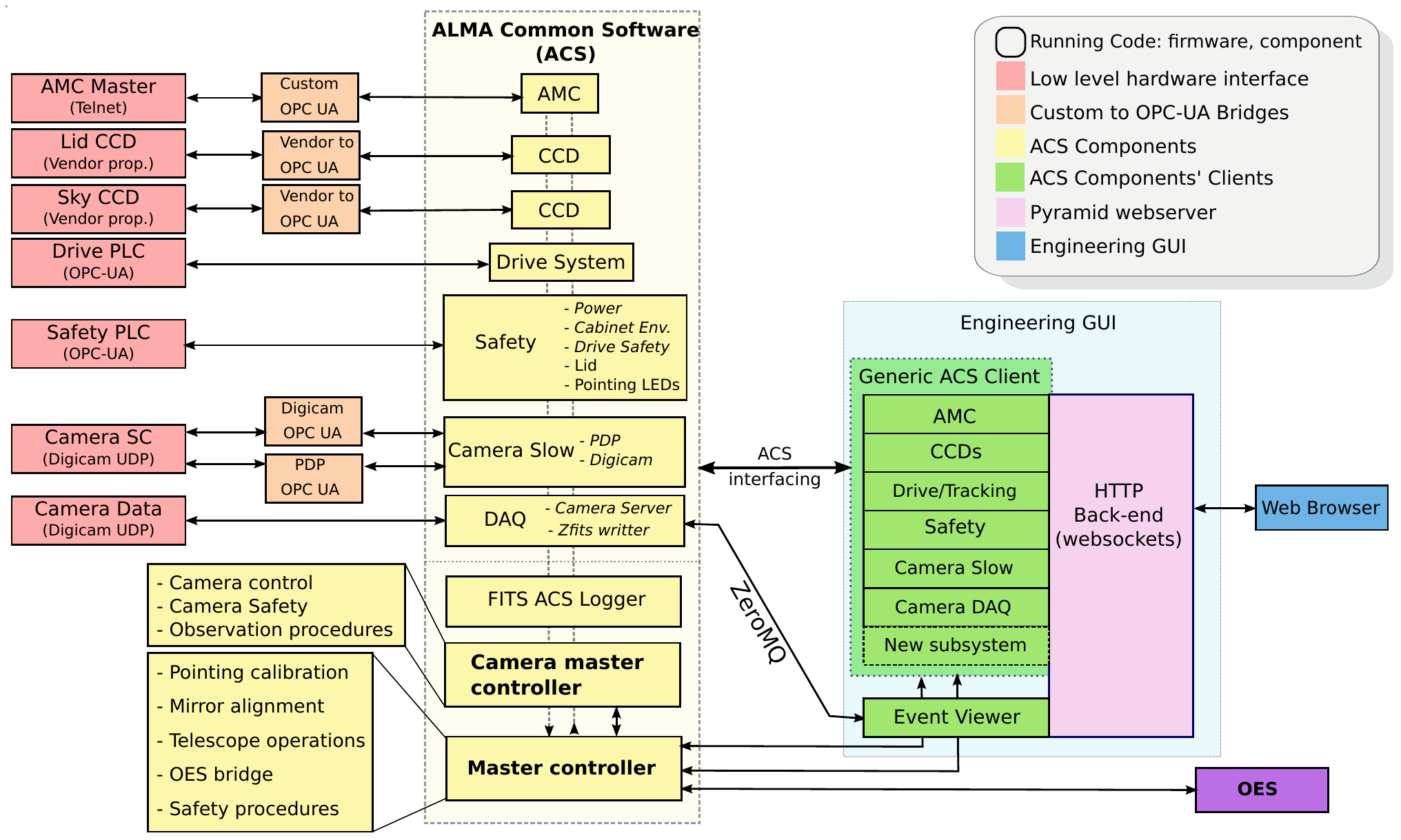}
    \caption{Diagram of the SST-1M control software. Taken from~\cite{Alispach:2024gvp}.}
    \label{fig:control}
\end{figure}

The SST-1M control system is built on the ALMA Common Software (ACS) framework, providing a modular and scalable architecture for managing all telescope subsystems. A web-based graphical user interface (GUI) allows for both local and remote operation, supporting multiple users and enabling real-time monitoring and control~\cite{ICRC25Operation}. A diagram of the control software and its interface with hardware and the GUI is show in Figure~\ref{fig:control}.


\subsection{The SiPM-based camera DigiCam}

The DigiCam camera is a fully digital, modular system designed specifically for the SST-1M \cite{CameraPaperHeller2017}. The camera features a photo-detection plane (PDP) composed of 1296 hexagonal silicon photomultipliers (SiPMs) pixels and light concentrator arranged in 108 modules. The SiPMs, developed in collaboration with Hamamatsu, are optimized for high photon detection efficiency and low noise. The front-end electronics include pre-amplifier and slow control boards that manage signal amplification, temperature compensation, and bias voltage regulation. Each pixel views an angular size of 0.24°, resulting in a total field of view of approximately 9°.
Signals from the SiPMs are digitized using flash analog-to-digital converters (FADCs) and processed by field-programmable gate arrays (FPGAs), which implement flexible triggering algorithms and data buffering. The system supports dead-time-free operation and high data throughput, with synchronization provided by the White Rabbit timing protocol \cite{ICRC25Operation}.



\section{Calibration}\label{sec:calib}

Accurate calibration is essential for the SST-1M telescopes to ensure reliable and consistent reconstruction of gamma-ray events. Wearouts of the telescope components such as the mirrors have to be monitored frequently.  Additionally, the night sky background (NSB) significantly affects the camera’s performance by inducing a voltage drop in the SiPMs, leading especially to reduced gain and photon detection efficiency. To maintain accurate photo-electron reconstruction, the impact of NSB is corrected during data processing, ensuring that the number of photoelectrons is properly recovered~\cite{Alispach:2024gvp}.

\subsection{Camera calibration parameter monitoring}

\begin{figure}
    \centering
    \includegraphics[width=\linewidth]{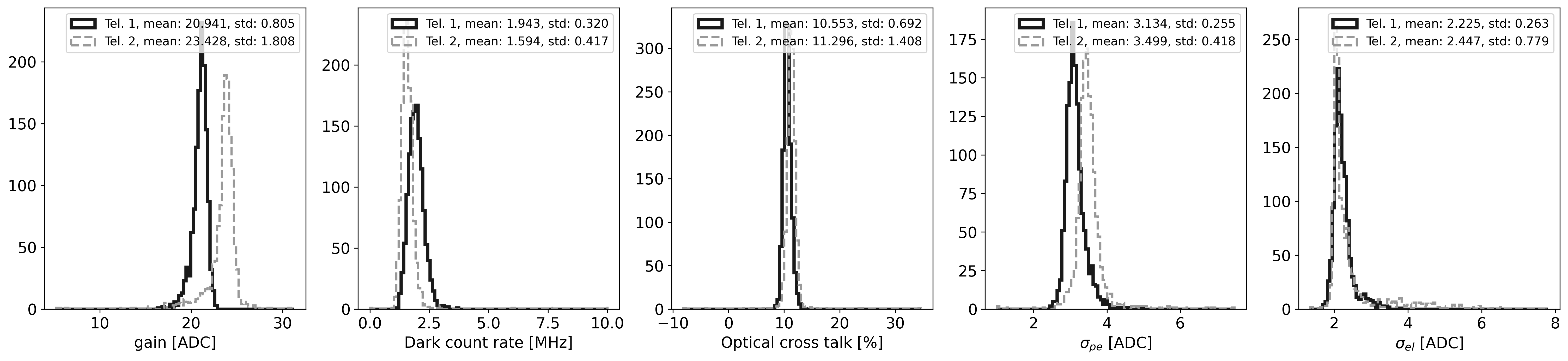}
    \caption{Calibration parameters measured for all pixels of the two telescope cameras SST-1M-1 (shown in dark) and SST-1M-2 (in gray) using dark count runs. From left to right, the panels display the gain, dark count rate (DCR), optical crosstalk, gain variation (smearing) $\sigma_{pe}$, and electronic noise $\sigma_e$ distributions. Taken from~\cite{Alispach:2024gvp}.}
    \label{fig:calib}
\end{figure}

Key camera parameters—such as gain, dark count rate, optical crosstalk, and electronic noise—are routinely monitored through dedicated dark runs conducted before and after each observation night. Measurements of these parameters across the camera are presented in Figure~\ref{fig:calib}. The narrow distribution of camera parameters ensures a uniform response to incoming light across all pixels. Furthermore, the stability of these parameters over time guarantees consistent and reliable measurements across multiple observation nights~\cite{CrabAA, Alispach:2024gvp}.

\subsection{Muon analysis}

\begin{figure}
    \centering
    \includegraphics[width=0.3\linewidth]{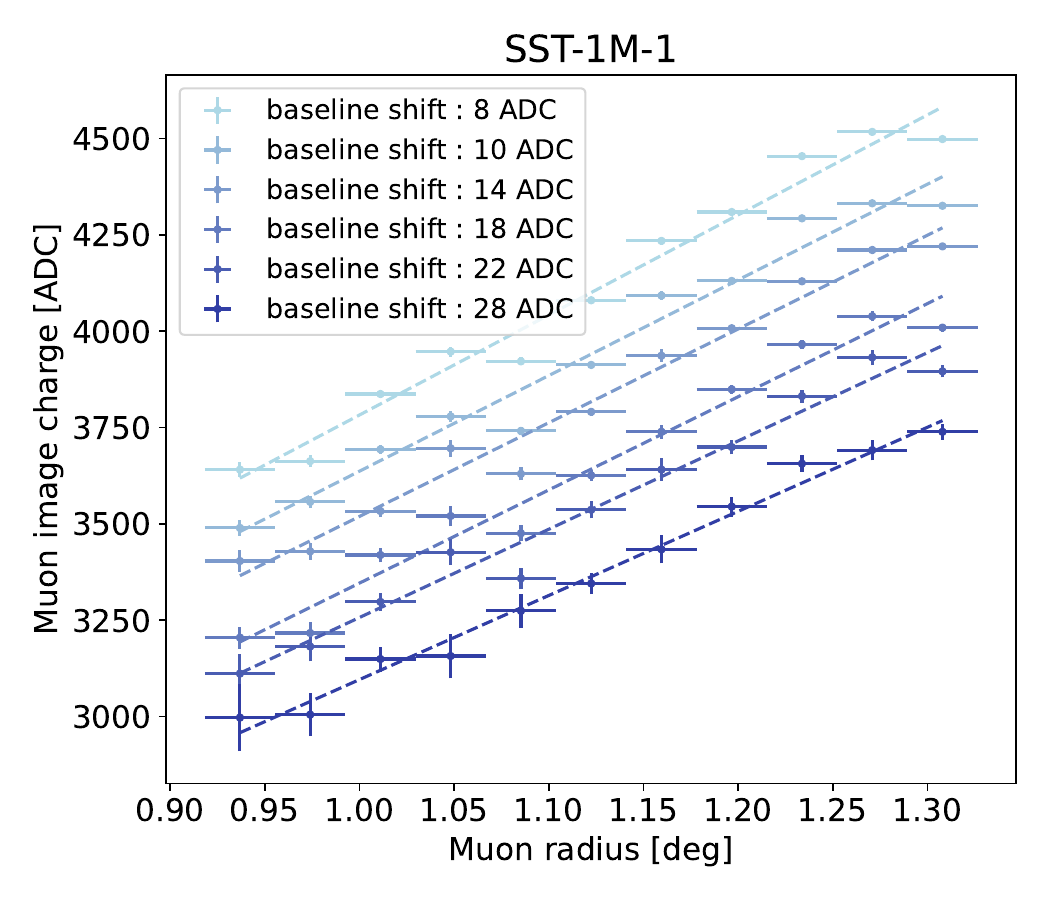}
    \includegraphics[width=0.3\linewidth]{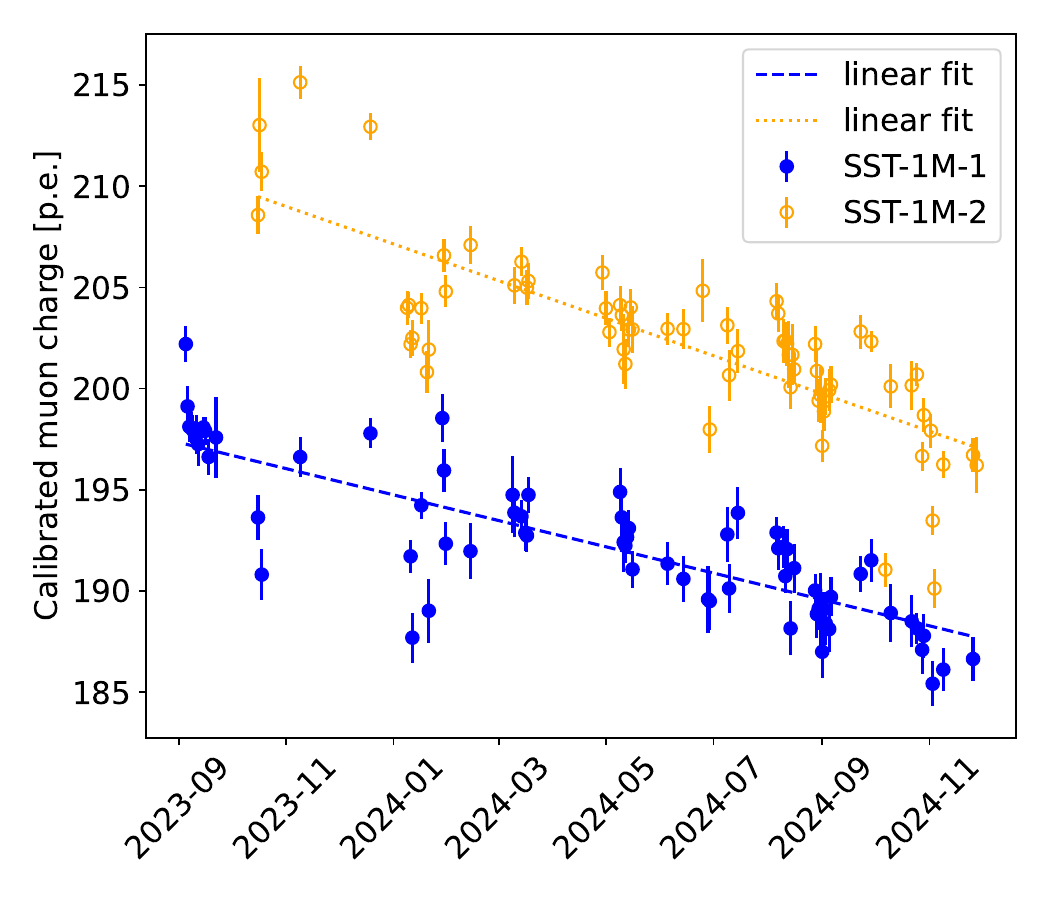}
    \caption{\textbf{Left:} Total charge measured (in ADC) in muon rings as a function of ring radius, shown for various baseline shifts corresponding to different NSB levels (for the SST-1M-1 telescope). \textbf{Right:} Total muon charge (in photoelectrons) recorded between September 2023 and November 2024 for SST-1M-1 (blue) and SST-1M-2 (orange). The values have been corrected to account for performance degradation due to NSB-induced effects on the camera. Taken from~\cite{CrabAA}.}
    \label{fig:muon}
\end{figure}

Muon rings provide a robust method for monitoring the optical efficiency of the SST-1M telescopes. When atmospheric muons pass near the telescope, they emit Cherenkov light that forms a circular image on the camera. By measuring the total charge in these rings as a function of the ring radius, one can assess the telescope’s light collection efficiency. Figure~\ref{fig:muon}-left shows the measured muon charge versus ring radius under various night sky background (NSB) conditions. After correcting for NSB-induced voltage drop effects, the temporal evolution of the optical efficiency can be extracted, as illustrated in Figure~\ref{fig:muon}-center. This approach enables the detection of performance degradation, such as the gradual decline in mirror reflectivity, which typically leads to a 2–5\% drop in optical throughput over several months. Muon rings thus serve as a stable and well-characterized calibration source, allowing for accurate reconstruction of the number of photoelectrons and correction of NSB-related biases.

\section{Observations}\label{sec:obs}

In this contribution, we highlight observations of several astrophysical gamma-ray sources conducted with the SST-1M telescopes, including the Crab Nebula, CTA 1, VER J2019+368, and Markarian 421. These sources span a range of astrophysical scenarios from pulsar wind nebulae to active galactic nuclei and illustrate the scientific potential of the SST-1M system in the multi-TeV regime. More detailed analyses are presented in accompanying contributions within these proceedings~\cite{ICRC25CTA1, ICRC25DragonFly, ICRC25Mrk421, ICRC25Crab} but are summarized in~\ref{sec:obs:others}. The differential flux sensitivity is shown in Figure~\ref{fig:crab}-right and detailed in accompanying contribution from these proceedings~\cite{ICRC25StereoPerf}.

The Crab Nebula with its well-characterized spectrum, below few tens of TeV, provides a benchmark for assessing the accuracy of the instrument response and reconstruction pipeline. All data presented in these proceedings were processed using the open-source analysis framework \href{https://github.com/SST-1M-collaboration/sst1mpipe}{\textit{sst1mpipe}}, developed specifically for the SST-1M telescopes. This pipeline ensures reproducibility and transparency in the data analysis and is publicly available to the broader community.

\subsection{Crab Nebula}\label{sec:obs:crab}

Between September 2023 and March 2024, the Crab Nebula was observed. These observations marked the first scientific campaign of the SST-1M system, aimed at validating its performance in both mono and stereo configurations and benchmarking its capabilities against established VHE gamma-ray observatories. The results of this observation campaign are briefly summarized here and were published recently in~\cite{CrabAA}. An up-to-date analysis is presented in these accompanying proceedings~\cite{ICRC25Crab} showing the Crab Nebula spectrum up to 100~TeV.

The data set included 33 hours of stereo data. Careful calibration of the telescopes using dark runs, muon ring analysis, and corrections for night sky background (NSB) effects were performed. Monte Carlo (MC) simulations were extensively used to model the instrument response and atmospheric conditions, and were tuned to match the observed data.

\begin{figure}
    \centering
    \includegraphics[width=0.3\linewidth]{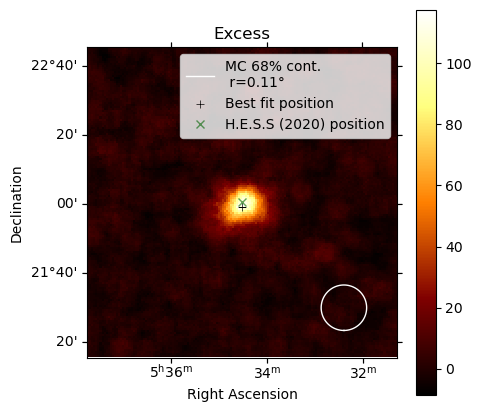}
    \includegraphics[width=0.3\linewidth]{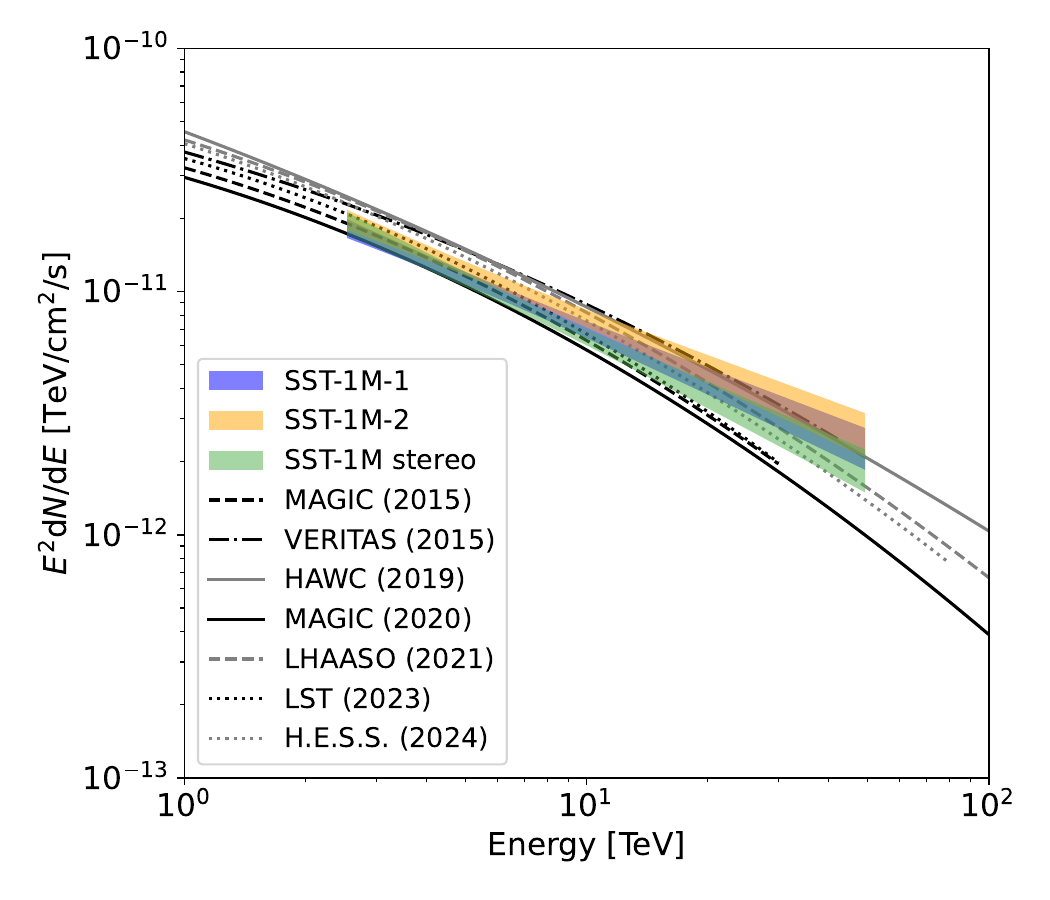}
    \includegraphics[width=0.3\linewidth]{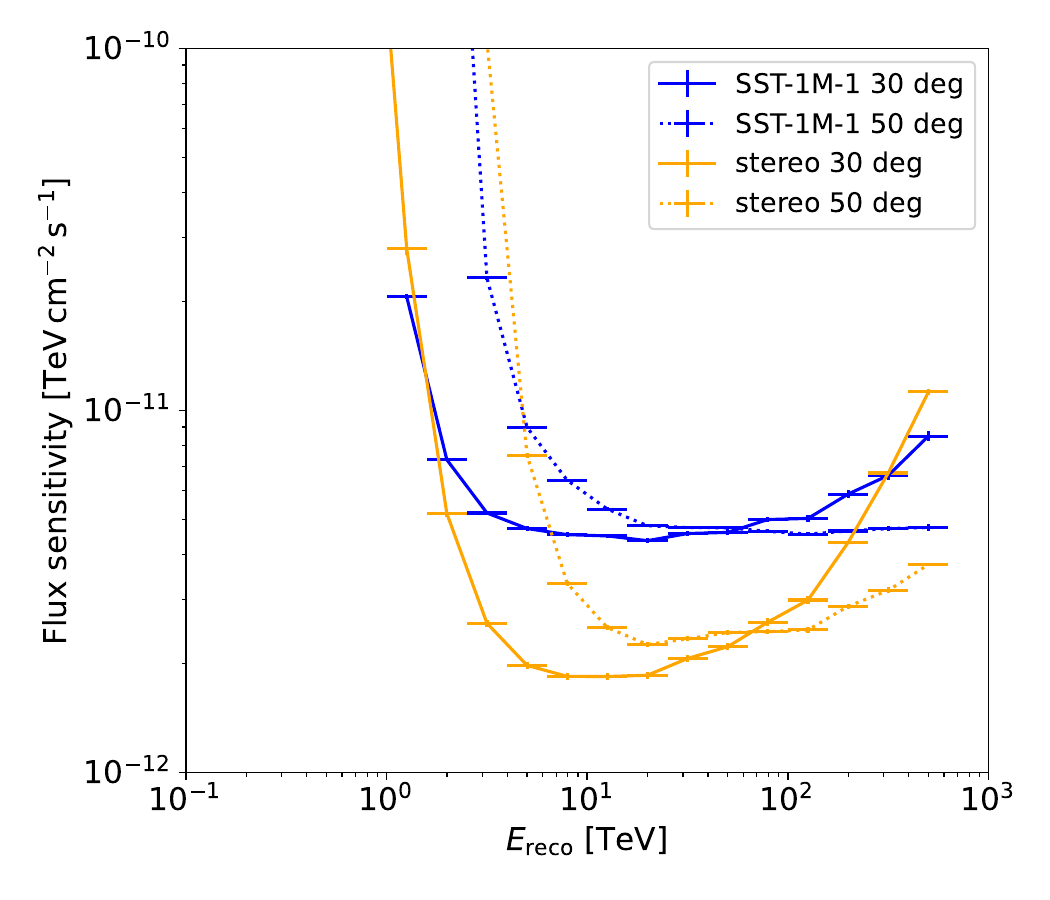}
    \caption{\textbf{Left:} Excess sky-map of the Crab Nebula observation. The plus "+" indicate the best fit source location while the cross "x", indicates the source coordinate found by the H.E.S.S. collaboration. \textbf{Center:} SED of the Crab Nebula measured with SST-1M telescopes. Different color bands represent the best-fitting spectra for the SST-1M-1 mono (blue), SST-1M-2 mono (yellow), and SST-1M stereo (green) datasets. The black lines represents measurement from other gamma-ray ground-based observatories. \textbf{Right:} Differential flux sensitivity as function of the reconstructed energy for a $5\sigma$ confidence level for the mono telescope system (blue) and the stereoscopic system (orange) evaluated at two zenith angles of 30° (dashed) and 50° (solid). Taken from~\cite{CrabAA}.}
    \label{fig:crab}
\end{figure}

Spectral analysis of the Crab Nebula was performed using a power-law model over the 2.5–50 TeV range. The results were consistent across both telescopes and the stereo system, with spectral indices $2.78\pm 0.10_{stat} \pm 0.08_{sys}$ (for the stereo) and flux normalizations in agreement with previous measurements from MAGIC, VERITAS, HAWC, and LHAASO (see Figure~\ref{fig:crab}-center). The reconstructed sky maps confirmed the source position within 0.02° of the known Crab coordinates (see Figure~\ref{fig:crab}-left), validating the pointing accuracy of the system. 

A measurement of baseline fluctuations in the pixel aligned with the Crab Nebula revealed a clear modulation correlated with the optical pulsations of the Crab pulsar. This observation confirms the timing precision of the SST-1M system (see~\cite{ICRC25Crab}).

\subsection{Galactic sources and AGNs}\label{sec:obs:others}

Since September 2023, the SST-1M telescopes have accumulated a total of 92.2 hours of Crab Nebula observations, covering zenith angles between 25° and 65°. The excellent agreement achieved in both spectral measurements and source localization confirms the readiness of the SST-1M system to enter a new phase of scientific operation. Recent improvements in the instrument response functions (IRFs), particularly across a range of zenith angles have significantly reduced systematic uncertainties~\cite{CrabAA}. These advancements now enable the SST-1M array to confidently begin targeted observations of additional gamma-ray sources, marking a major step forward in its scientific capabilities.

\begin{figure}
    \centering
    \includegraphics[width=0.3\linewidth]{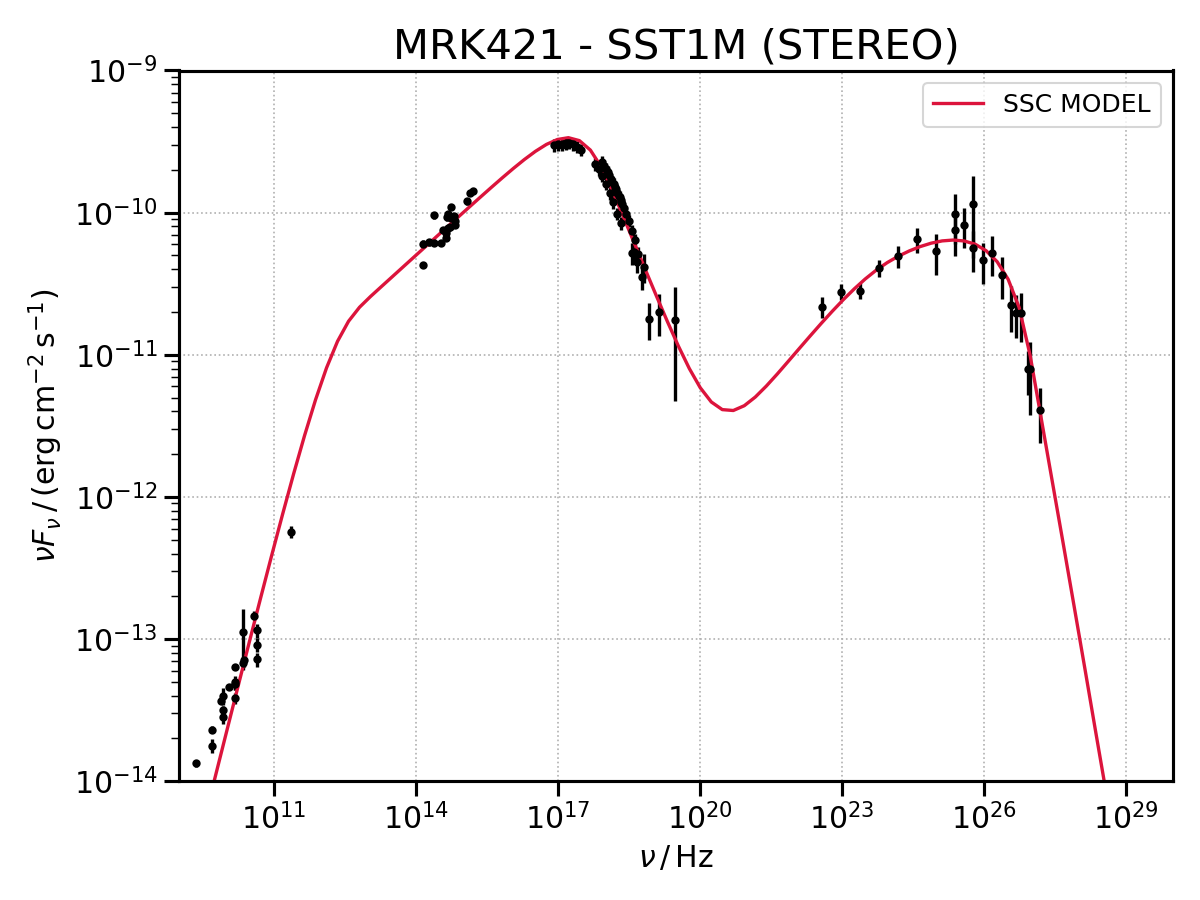}
    \includegraphics[width=0.25\linewidth]{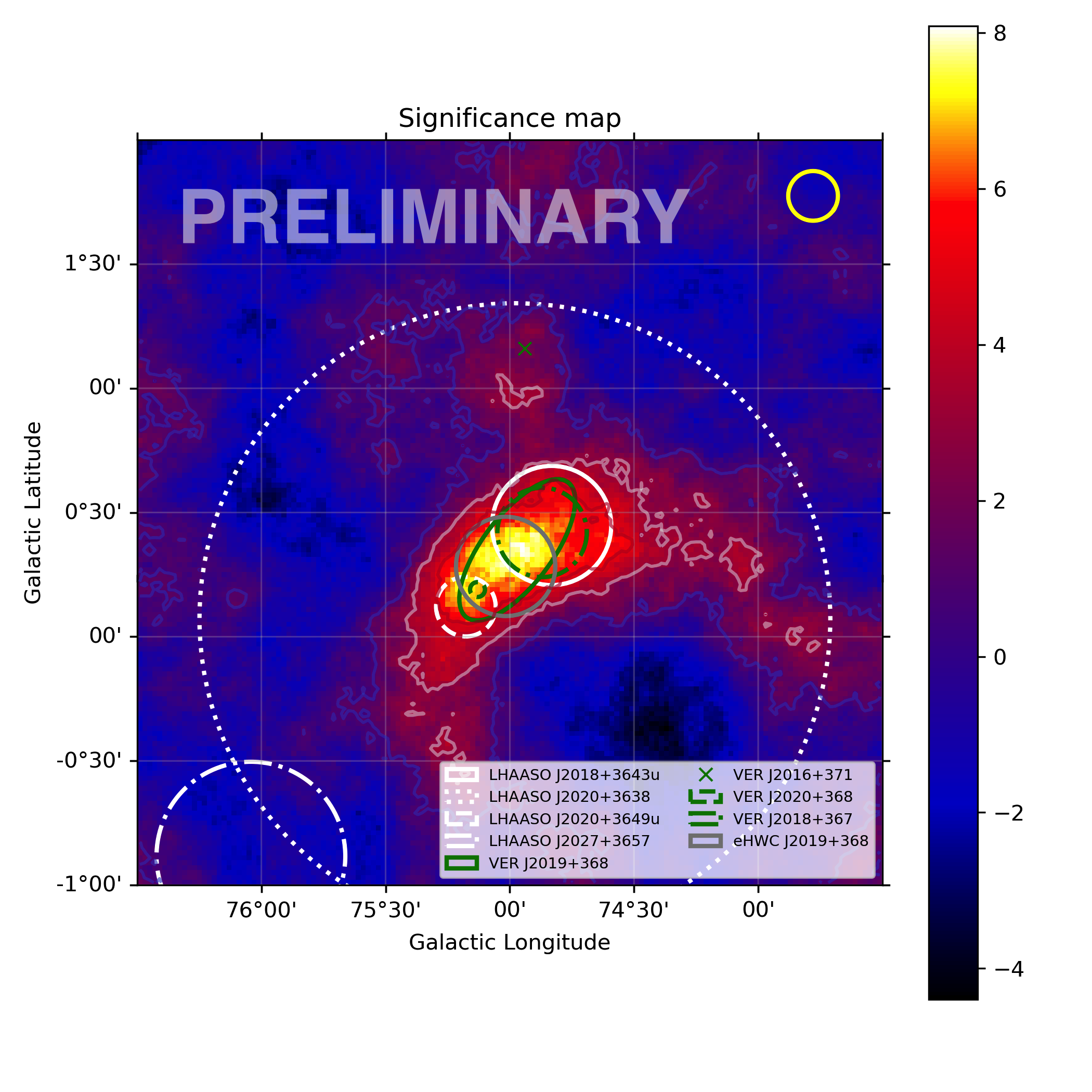}
    \includegraphics[width=0.3\linewidth]{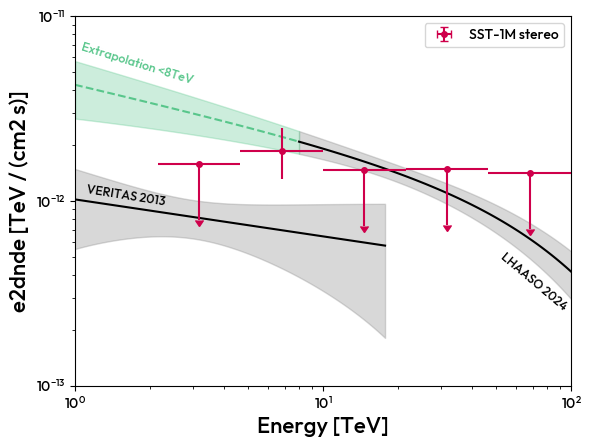}
    \caption{\textbf{Left:} SCC model of Mkr 421 with measurements of the SST-1M stereoscopic system .Taken from~\cite{ICRC25Mrk421}. \textbf{Center:} Skymap of local significance for the VER J2019+368 region smeared with a gaussian kernel of 0.25°. The VHE sources coordinates in the region are indicated. Taken from~\cite{ICRC25DragonFly}. \textbf{Right:} Upper limits of the spectrum of CTA 1 measured by the SST-1M telescopes compared to the spectra of VERITAS an LHAASO. Taken from~\cite{ICRC25CTA1}.}
    \label{fig:obs:others}
\end{figure}

For instance, observations of the blazar Markarian 421 (Mrk 421) have been conducted. Using data collected above 0.5 TeV, the parameters of a synchrotron self-Compton (SSC) model were derived, demonstrating consistency with results from other experiments such as HAWC. The SSC model obtained with the SST-1M gamma-ray (combined with other observatories at different wavelength) data is shown in Figure~\ref{fig:obs:others}-left. A detailed description of this analysis is provided in these proceedings~\cite{ICRC25Mrk421}. Additionally, the SST-1M stereoscopic system issued its first Astronomer’s Telegram (ATel \#16533), reporting a significant increase in gamma-ray flux from Mrk 421 above 2 TeV, observed on March 15, 2024.

Observations of the extended region Dragonfly lead to detection of the main emission VHE region resolving VERJ2019+368 and VER J2016+371 components. These observations contribute to the ongoing evaluation of the SST-1M telescopes’ off-axis performance, which is crucial for wide field-of-view analyses. A preliminary sky-map of the VER 2019+368 region is presented in Figure~\ref{fig:obs:others}-center. Further details and results of this analysis are discussed in these proceedings~\cite{ICRC25DragonFly}.

A total of 30 hours of observations were conducted on CTA 1 using the SST-1M telescopes. These data yielded a detection significance of $3.5\sigma$, suggesting a potential signal that could be confirmed with additional observation time in stereo mode. The observed excess is spatially offset by 0.25° with the pulsar PSR J0007+7303. Although a firm $5\sigma$ detection has not yet been achieved, upper limits on the gamma-ray spectrum have been derived and are presented in Figure~\ref{fig:obs:others}-right. This analysis is discussed in greater detail in these proceedings~\cite{ICRC25CTA1}.

\section{The future of SST-1M}\label{sec:future}

Overall, the Crab Nebula and other gamma-ray sources observations confirmed that the SST-1M system performs reliably and accurately in both mono and stereo modes. The results validate the instrument model and analysis pipeline, demonstrating that SST-1Ms are well-suited for multi-TeV gamma-ray astronomy, including the study of extended sources and transient phenomena.

The SST-1M telescope array deployment at a higher altitude site is being investigated. Two promising candidate locations are currently under consideration: the Indian Astronomical Observatory and the Pierre Auger Observatory. Both sites offer significant advantages in terms of atmospheric transparency and existing infrastructure. The performances of the SST-1M at these location is compared to the current site in these proceedings~\cite{ICRC25StereoPerf}.

Ongoing simulation campaigns are assessing the performance of the SST-1M array at prospective deployment sites, including optimization of telescope layouts and potential integration with existing instruments such as SWGO. A recent study (see these proceedings~\cite{ICRC25Hybrid}) explores a hybrid detection approach, combining SST-1M telescope images with gamma/hadron discrimination parameters: the \textit{LCm} and $P_\alpha^{\text{tail}}$—from SWGO’s Water Cherenkov Tanks. This combined analysis demonstrates a 30\% improvement in sensitivity above 10 TeV, driven by enhanced background rejection enabled by the inclusion of the SWGO parameters.

These efforts aim to identify the most effective stereo and hybrid configurations for maximizing sensitivity and sky coverage. In particular, efforts are being done to optimize the array layout using differential programming for end-to-end optimization aiming at building a generic optimization pipeline for future Cherenkov telescopes observatories.

\section*{Acknowledgments}

\setlength{\baselineskip}{10pt}
{\scriptsize This publication was created as part of the projects funded in Poland by the Minister of Science based on agreements number 2024/WK/03 and DIR/\-WK/2017/12. The construction, calibration, software control and support for operation of the SST-1M cameras is supported by SNF (grants CRSII2\_141877, 20FL21\_154221, CRSII2\_160830, \_166913, 200021-231799), by the Boninchi Foundation and by the Université de Genève, Faculté de Sciences, Département de Physique Nucléaire et Corpusculaire. The Czech partner institutions acknowledge support of the infrastructure and research projects by Ministry of Education, Youth and Sports of the Czech Republic (MEYS) and the European Union funds (EU), MEYS LM2023047, EU/MEYS CZ.02.01.01/00/22\_008/0004632, CZ.02.01.01/00/22\_010/0008598, Co-funded by the European Union (Physics for Future – Grant Agreement No. 101081515), and Czech Science Foundation, GACR 23-05827S.}

\bibliographystyle{JHEP}
{\scriptsize
\bibliography{bibliography.bib}

\providecommand{\href}[2]{#2}\begingroup\raggedright\begin{thebibliography}{10}

\bibitem{Alispach:2024gvp}
C.~Alispach et~al., \emph{{The SST-1M imaging atmospheric Cherenkov telescope for gamma-ray astrophysics}}, \href{https://doi.org/10.1088/1475-7516/2025/02/047}{\emph{JCAP} {\bfseries 02} (2025) 047}.

\bibitem{ICRC25Operation}
D.~Mandat et~al., \emph{{Operation of the SST-1M Cherenkov telescope gamma ray stereoscopic system}}, {\emph{PoS} {\bfseries ICRC2025} (2026) }.

\bibitem{CrabAA}
{Alispach, C.} et~al., \emph{Observation of the crab nebula with the single-mirror small-size telescope stereoscopic system at low altitude}, \href{https://doi.org/10.1051/0004-6361/202555292}{\emph{A\&A} {\bfseries 699} (2025) A255}.

\bibitem{ICRC25CTA1}
B.~Lacave et~al., \emph{{SST-1M Observation of CTA 1}}, {\emph{PoS} {\bfseries ICRC2025} (2026) }.

\bibitem{ICRC25DragonFly}
J.~Jury\v{s}ek et~al., \emph{{Observation of VER 2019+368 with the SST-1M stereoscopic system}}, {\emph{PoS} {\bfseries ICRC2025} (2026) }.

\bibitem{ICRC25Mrk421}
M.S.~Reddy et~al., \emph{{SST-1M observation of Markarian 421}}, {\emph{PoS} {\bfseries ICRC2025} (2026) }.

\bibitem{ICRC25Crab}
T.~Tavernier et~al., \emph{{Calibration and Performance Validation of the SST-1M Telescopes Using Crab Nebula Observations}}, {\emph{PoS} {\bfseries ICRC2025} (2026) }.

\bibitem{ICRC25Hybrid}
A.~Bakalova et~al., \emph{{Hybrid concept of detection for a wide-field gamma-ray observatory using Cherenkov telescopes}}, {\emph{PoS} {\bfseries ICRC2025} (2026) }.

\bibitem{CameraPaperHeller2017}
M.~Heller et~al., \emph{{\it An innovative silicon photomultiplier digitizing camera for gamma-ray astronomy}}, \href{https://doi.org/10.1140/epjc/s10052-017-4609-z}{\emph{EPJ C} {\bfseries 77} (2017) 47}.

\bibitem{ICRC25StereoPerf}
P.~Cechvala et~al., \emph{{Stereo performance of SST-1M at different altitudes}}, {\emph{PoS} {\bfseries ICRC2025} (2026) }.

\end{thebibliography}\endgroup
}
\clearpage
\section*{Full Authors List: SST-1M Collaboration}
\scriptsize
\noindent
C.~Alispach$^1$,
A.~Araudo$^2$,
M.~Balbo$^1$,
V.~Beshley$^3$,
J.~Bla\v{z}ek$^2$,
J.~Borkowski$^4$,
S.~Boula$^5$,
T.~Bulik$^6$,
F.~Cadoux$^`$,
S.~Casanova$^5$,
A.~Christov$^2$,
J.~Chudoba$^2$,
L.~Chytka$^7$,
P.~\v{C}echvala$^2$,
P.~D\v{e}dic$^2$,
D.~della Volpe$^1$,
Y.~Favre$^1$,
M.~Garczarczyk$^8$,
L.~Gibaud$^9$,
T.~Gieras$^5$,
E.~G{\l}owacki$^9$,
P.~Hamal$^7$,
M.~Heller$^1$,
M.~Hrabovsk\'y$^7$,
P.~Jane\v{c}ek$^2$,
M.~Jel\'inek$^{10}$,
V.~J\'ilek$^7$,
J.~Jury\v{s}ek$^2$,
V.~Karas$^{11}$,
B.~Lacave$^1$,
E.~Lyard$^{12}$,
E.~Mach$^5$,
D.~Mand\'at$^2$,
W.~Marek$^5$,
S.~Michal$^7$,
J.~Micha{\l}owski$^5$,
M.~Miro\'n$^9$,
R.~Moderski$^4$,
T.~Montaruli$^1$,
A.~Muraczewski$^4$,
S.~R.~Muthyala$^2$,
A.~L.~Müller$^2$,
A.~Nagai$^1$,
K.~Nalewajski$^5$,
D.~Neise$^{13}$,
J.~Niemiec$^5$,
M.~Niko{\l}ajuk$^9$,
V.~Novotn\'y$^{2,14}$,
M.~Ostrowski$^{15}$,
M.~Palatka$^2$,
M.~Pech$^2$,
M.~Prouza$^2$,
P.~Schovanek$^2$,
V.~Sliusar$^{12}$,
{\L}.~Stawarz$^{15}$,
R.~Sternberger$^8$,
M.~Stodulska$^1$,
J.~\'{S}wierblewski$^5$,
P.~\'{S}wierk$^5$,
J.~\v{S}trobl$^{10}$,
T.~Tavernier$^2$,
P.~Tr\'avn\'i\v{c}ek$^2$,
I.~Troyano Pujadas$^1$,
J.~V\'icha$^2$,
R.~Walter$^{12}$,
K.~Zi{\c e}tara$^{15}$ \\

\noindent
$^1$D\'epartement de Physique Nucl\'eaire, Facult\'e de Sciences, Universit\'e de Gen\`eve, 24 Quai Ernest Ansermet, CH-1205 Gen\`eve, Switzerland.
$^2$FZU - Institute of Physics of the Czech Academy of Sciences, Na Slovance 1999/2, Prague 8, Czech Republic.
$^3$Pidstryhach Institute for Applied Problems of Mechanics and Mathematics, National Academy of Sciences of Ukraine, 3-b Naukova St., 79060, Lviv, Ukraine.
$^4$Nicolaus Copernicus Astronomical Center, Polish Academy of Sciences, ul. Bartycka 18, 00-716 Warsaw, Poland.
$^5$Institute of Nuclear Physics, Polish Academy of Sciences, PL-31342 Krakow, Poland.
$^6$Astronomical Observatory, University of Warsaw, Al. Ujazdowskie 4, 00-478 Warsaw, Poland.
$^7$Palack\'y University Olomouc, Faculty of Science, 17. listopadu 50, Olomouc, Czech Republic.
$^8$Deutsches Elektronen-Synchrotron (DESY) Platanenallee 6, D-15738 Zeuthen, Germany.
$^9$Faculty of Physics, University of Bia{\l}ystok, ul. K. Cio{\l}kowskiego 1L, 15-245 Bia{\l}ystok, Poland.
$^{10}$Astronomical Institute of the Czech Academy of Sciences, Fri\v{c}ova~298, CZ-25165 Ond\v{r}ejov, Czech Republic.
$^{11}$Astronomical Institute of the Czech Academy of Sciences, Bo\v{c}n\'i~II 1401, CZ-14100 Prague, Czech Republic.
$^{12}$D\'epartement d'Astronomie, Facult\'e de Science, Universit\'e de Gen\`eve, Chemin d'Ecogia 16, CH-1290 Versoix, Switzerland.
$^{13}$ETH Zurich, Institute for Particle Physics and Astrophysics, Otto-Stern-Weg 5, 8093 Zurich, Switzerland.
$^{14}$Institute of Particle and Nuclear Physics, Faculty of Mathematics and Physics, Charles University, V Hole\v sovi\v ck\' ach 2, Prague 8, Czech~Republic.
$^{15}$Astronomical Observatory, Jagiellonian University, ul. Orla 171, 30-244 Krakow, Poland.

\end{document}